\documentclass[prl,aps,twocolumn,amsmath,amssymb,showpacs]{revtex4}

\usepackage{graphicx}
\usepackage{color}
\usepackage[dvips]{epsfig}

\usepackage{soul}
\setstcolor{red}

\newcommand{\be}{\begin{equation}}
\newcommand{\ee}{\end{equation}}

\newcommand{\D}{\hat \Psi}

\renewcommand{\vec}[1]{\mathbf{#1}}

\begin{document}

\title{Bose-Einstein condensation of stationary-light polaritons}
\author{Michael Fleischhauer, Johannes Otterbach, and Razmik G. Unanyan}
\affiliation{Fachbereich Physik, Technische Universit\"{a}t
Kaiserslautern, D-67663 Kaiserslautern,
Germany}

\begin{abstract}
We propose and analyze a mechanism for Bose-Einstein condensation of stationary dark-state polaritons.
Dark-state polaritons (DSPs) are formed in the interaction of light with laser-driven 
3-level $\Lambda$-type atoms and are the basis of phenomena such as electromagnetically induced transparency (EIT),
ultra-slow and stored light. They have long intrinsic lifetimes and in a stationary set-up with two
counterpropagating control fields of equal intensity have a 3D quadratic dispersion profile with variable
effective mass. Since DSPs are bosons they can undergo a Bose-Einstein condensation at a
critical temperature which can be many orders of magnitude larger than that of atoms. We show that thermalization
of polaritons can occur via elastic collisions mediated by a resonantly enhanced
optical Kerr nonlinearity on a time scale short compared to the decay time. Finally condensation can
be observed by turning stationary into propagating polaritons and monitoring the emitted light. 
\end{abstract}

\date{\today}

\maketitle

Since the critical temperature of Bose-Einstein condensation (BEC) \cite{Bose,Einstein} is inversely proportional to
the mass of the particles already early on in the history of the field quasi-particles where considered as candidates
for condensation at high temperatures \cite{Blatt-PR-1962}. 
Very recently this subject regained a lot
of attention triggered by experimental breakthroughs in microcavity exciton-polaritons \cite{Littlewood-Science-2007,Kasprzak-Nature-2006,Balili-Nature-2007}
and thin-film magnons \cite{Demokritov-Nature-2006}. 
Yet major drawbacks of these systems are the limitation to quasi-condensation
in two dimensions, and in the case of excitons, the necessity of a cavity which makes 
a distinction from lasing difficult \cite{Snoke-Nature-2006}.
We here propose a mechanism for true condensation in three spatial dimensions
employing a different kind of polaritonic quasi-particles, called dark-state polaritons (DSP) \cite{Zimmer-PRA-2008}. 
DSPs \cite{Fleischhauer-PRL-2000} emerge in the Raman interaction of lasers with 3-level quantum systems 
and are the basis of ultra-slow \cite{Hau-PRL-1999}, stopped \cite{Phillips-PRL-2001,Liu-Nature-2001}, and stationary light \cite{Bajcsy-Nature-2003}.
They have a considerably longer lifetime than exciton polaritons or magnons, provide a 3D quadratic dispersion with variable mass, and have very high condensation temperatures. DSPs can easily be created and thermalization can be 
achieved on a time scale much shorter than their lifetime.
Finally condensation can easily be observed by transforming stationary DSPs into light pulses.

If an optically thick ensemble of three-level quantum systems is irradiated by a strong coherent
coupling laser it can become transparent for a probe field within a certain frequency range close to the
two-photon Raman resonance.  
Associated with this phenomenon known as electromagnetically induced transparency (EIT) \cite{Harris-Physics-Today-1997,Fleischhauer-RMP-2005} is the formation of polaritonic eigensolutions, called 
dark-state polaritons (DSP) \cite{Fleischhauer-PRL-2000}, which are superpositions of the probe-field amplitude
and a collective Raman excitation. The mixing angle that determines the admixtures of field and matter degrees
of freedom can be changed by varying the strength of the control laser.
At the transparency frequency the dispersion of dark-state polaritons is 
linear with a slope determined by the mixing angle. This situation changes if two counter-propagating 
control laser  with comparable intensities are used rather than one. The two control
laser lead to a quasi-stationary probe-field pattern, known as stationary light \cite{Bajcsy-Nature-2003,Zimmer-OptComm-2006}.
We show that the dark polaritons of stationary light 
\cite{Zimmer-PRA-2008} behave like massive particles in 3D
and can undergo Bose-condensation at hight temperatures.

\begin{figure}[hbt]
	\centering
   \epsfig{file=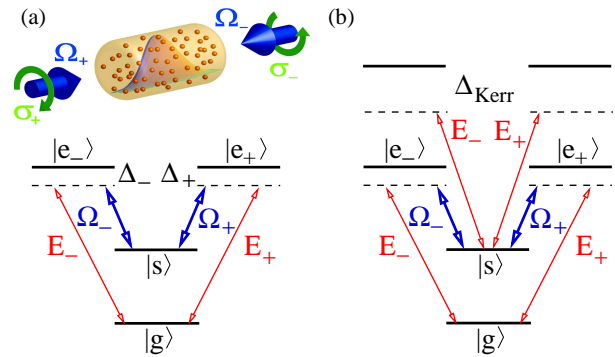, width=8 cm}
      \caption{(Color online) a) The Raman interaction of two counterpropagating control lasers of Rabi-frequencies
$\Omega_\pm$  and opposite circular polarization coupling to the $|s\rangle - |e_\pm\rangle$ transitions
of a double $\Lambda$ system generates a quasi-stationary pattern of Stokes fields $E_\pm$, called
stationary light. The stationary dark-state polaritons formed in this interaction do not decay radiatively
and have a quadratic dispersion profile in all directions. b) Off resonant coupling of the Stokes fields
$E_\pm$ with other excited states leads to ac-Stark shift of lower level $|s\rangle$ resulting in a resonantly enhanced Kerr-type 
self-interaction of $E_\pm$.
}
      \label{fig1}
\end{figure}
We here consider a generalization of the original stationary-light scheme 
\cite{Bajcsy-Nature-2003}, which is shown in Fig.\ref{fig1}a). It involves two parallel $\Lambda$ transitions 
consisting of the common ground levels $|g\rangle$ and $|s\rangle$ and the excited states
$|e_\pm\rangle$, coupled to opposite circular polarizations of quantized probe fields $\hat E_\pm$ and control lasers described by the Rabi-frequencies $\Omega_\pm$ \cite{Zimmer-PRA-2008}. Both $\Lambda$ schemes are in two-photon resonance which guarantees EIT. For the present discussion we will assume that the control fields are homogeneous and constant in time and thus set $\Omega_\pm =\Omega_\pm^*$. The advantage of this set-up as compared to the original one, where both
counterpropagating control laser couple to the same dipole transition is that in the present scheme higher
spatial harmonics of the ground-state coherence do not occur and that there is no point in space where the
total intensity of the control laser vanishes. 
The single-photon detunings of the upper states are denoted as $\Delta_+$ and $\Delta_-$ respectively. The fields
$\hat E_+$ $(\hat E_-)$ and $\Omega_+$ ($\Omega_-)$ propagate in the $+z$ $(-z)$ direction with 
wavenumbers $k_p$  ($-k_p$), and $k_c$ ($-k_c$).

We introduce normalized field amplitudes $\hat{\cal E}_\pm$ that vary slowly in space and time via $\hat E_\pm(\vec r,t) = \sqrt{\frac{\hbar\omega}{2\varepsilon_0}} \left(\hat {\cal E}_\pm(\vec r,t) \exp\{-i(\omega_p t \mp k_pz)\} + h.a.\right)$, and continuous atomic-flip operators $\hat\sigma_{\mu\nu}(\vec r,t) = \frac{1}{\Delta N}\sum_{j\in \Delta V(\vec r)} \hat \sigma^j_{\mu\nu}$, with $\hat\sigma_{\mu\nu}^j
\equiv |\mu\rangle_{jj}\langle \nu|$ being the flip operator of the $j$th atom.
The sum is taken over a small volume $\Delta V$ around $\vec r$ containing $\Delta N$ atoms.

\begin{figure}[hbt]
	\centering
     \epsfig{file=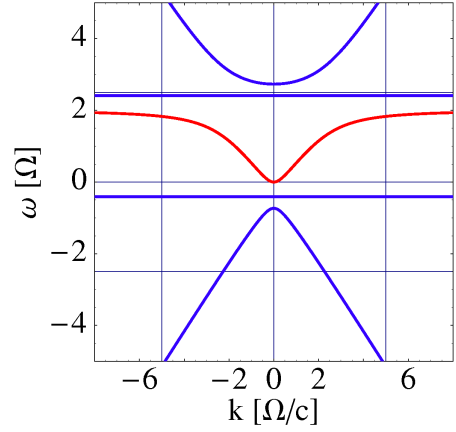, width=6cm}
      \caption{(Color online) Dispersion of stationary dark-state polariton (red) and other quasi-particle
solutions (blue) in propagation direction for equal detunings and control field Rabi-frequencies
$\Delta_+=\Delta_-=\Delta$, and $\Omega_+=\Omega_-=\Omega$, $\tan\theta = \tan\phi =1$, $\Delta/\Omega = 2$. Since only the real parts of the eigenvalues are of interest here, we set $\gamma=0$.
One clearly recognizes a quadratic profile for the dark polariton around $k=0$. It should be noted that
the polariton is defined by the slowly varying envelopes of field and matter variables and thus the
wavenumber $k$ refers to spatial modulations of the slowly varying amplitudes. 
}
      \label{fig2}
\end{figure}

Assuming that in the absence of the probe fields the control laser pump all population in the ground state $|g\rangle$,
the dynamical equations read in the linear response limit, i.e. for a small probe intensity
\begin{eqnarray}
\frac{\partial}{\partial t}\hat \sigma_{gs} &=& i\Omega_+ \hat\sigma_{ge_+} +i\Omega_-\hat\sigma_{ge_-},
\label{eq:sigma-gs}\\
\frac{\partial}{\partial t}\hat\sigma_{ge_\pm} &=& -\Gamma_{ge_\pm}\hat\sigma_{ge_\pm}+i\Omega_\pm\hat\sigma_{gs}
+i g\sqrt{n}\, \hat {\cal E}_\pm,\label{eq:sigma-ge}\\
\Bigl[\frac{\partial}{\partial t}\pm c\frac{\partial}{\partial z}&-& i \frac{c}{2 k_p}
\Delta\Bigr]\hat {\cal E}_\pm = ig\sqrt{n}\hat \sigma_{ge_\pm}.
\end{eqnarray}
Here $n$ is the atom density and $g=\frac{\wp}{\hbar}\sqrt{\frac{\hbar \omega}
{2\varepsilon_0}}$ is the common coupling constant of both probe fields with $\wp$ denoting the
dipole matrix element. $\Gamma_{\pm}=\gamma+i\Delta_{\pm}$, where $\gamma$ is the transverse decay rate of the transitions $|e_\pm\rangle - |g\rangle$ and $k_p=\omega_p/c$ is the carrier wavenumber of the probe field. If the characteristic time- and length-scales of the probe fields $T$ and $L$ are sufficiently large, such that the adiabaticity conditions $T \ll \left\{L_{\rm abs}/c, \gamma^{-1}\right\}$ and $L\gg L_{\rm abs}$
hold, where $L_{abs}\equiv \gamma c/(g^2 n)$ defines the resonant absorption length, the 
system can be described by polariton-like quasi-particles 
\cite{Zimmer-PRA-2008}. One of these solutions, called dark-state polariton (DSP) 
does not involve excited states and is thus immune to spontaneous emission. 
\begin{equation}
\hat \Psi(\vec r,t) = \cos\theta \left(\cos\phi \hat {\cal E}_+(\vec r,t) +\sin\phi \hat 
{\cal E}_-(\vec r,t)\right) -\sin\theta \hat S(\vec r,t)
\end{equation}
where $\hat S\equiv \sqrt{n} \hat \sigma_{gs}(\vec r,t)$ and the mixing angles are 
defined by $\tan^2\theta = g^2 n/\Omega^2$ with $\Omega^2 
=\Omega_+^2+\Omega_-^2$, and $\tan^2\phi =\Omega_- ^2/\Omega_+ ^2$. 
The longitudinal dispersion relation of the stationary DSPs is plotted in Fig.\ref{fig2} for $\tan\phi=1$
(red line). Also shown are the energies of the other 
eigensolutions. It proves useful for calculations to introduce special superpositions of these eigensolution termed bright-state polaritons.
\begin{align}
\hat\Phi_1(\vec r,t)\,&=\,-\sin\phi\hat{E}_+(\vec r,t)+\cos\phi\hat{E}_-(\vec r,t), \\
\hat\Phi_2(\vec r,t)\,&=\,-\sin\theta\big(\cos\phi\hat{E}_+(\vec r,t) 
+\sin\phi\hat{E}_-(\vec r,t)\big)\nonumber \\  &\qquad -\cos\theta\,\hat S(\vec r,t).
\end{align}

It is easy to derive the equation of motion of the DSP up to second order including the transverse directions. 
One finds for $\Delta_\pm=\Delta \gg \gamma$ and $\phi=\pi/4$
\begin{eqnarray}
\left[\frac{\partial}{\partial t} - i v_{\rm gr} L_{\rm abs} \left(i +\frac{\Delta}{\gamma}\right)\frac{\partial ^2}{\partial z^2} - i \frac{v_{\rm gr}}{2 k_p} \Delta_\perp\right]\hat \Psi(\vec r,t)=0,\label{eq:Psi}
\end{eqnarray}
where $v_{\rm gr}= c \cos^2 \theta$ is the group velocity of the polariton if only one of the
two coupling fields is present.
As opposed to microcavity exciton-polaritons, stationary DSPs behave as free, 
massive Schr\"odinger particles in all three spatial directions with a tensorial mass,
$m_\perp = m v_{\rm rec}/v_{\rm gr}$, 
$m_\parallel^{-1}  = m_\perp^{-1} \,
2k_pL_{\rm abs}\left({\Delta}/{\gamma}+i\right)$.
Here we have introduced the recoil velocity of the atom at the probe frequency
$v_{\rm rec} = \hbar k_p/m$ and the mass of the atoms $m$ as comparative scales. 
The effective mass  can be varied via the strength of the control laser, i.e. via the mixing angle $\theta$,
and is typically several orders of magnitude smaller than the
mass of the atoms. 
The longitudinal mass $m_\parallel$ is again smaller than the transverse 
mass by the ratio of resonant wavelength to absorption length and the inverse normalized detuning $\gamma/\Delta$. It
also has a small imaginary component, which describes the absorption of high-$k$ components 
well known from EIT \cite{Fleischhauer-RMP-2005}.
As can be seen from Fig.\ref{fig2} the quadratic dispersion of the dark-state polariton turns over 
into a linear dispersion at higher frequencies and eventually saturates giving rise to a finite
band of frequencies whose width is of the order of the single-photon detuning $\Delta$ 
for $|\Delta| \gg \Omega,\gamma$. 

Using the well known expression for the critical temperature of condensation for a homogeneous, ideal gas 
and expressing the polariton mass in terms of the mass of the atoms one finds
\begin{equation}
T_c = T_c^{\rm atom} \left(\frac{\rho_{\textrm{DSP}}}{n}\right)^{2/3} \frac{v_{\rm gr}}{v_{\rm rec}} \left(2 k_p L_{\rm abs} \frac{\Delta}{\gamma}\right)^{1/3}.
\end{equation}
Here we have introduced the critical temperature of the atoms $k_B T_c^{\rm atom} = 2\pi n^{2/3}/[\zeta(3/2)^{2/3} m]$
as a comparative scale. $\rho_{\rm DSP}$ is the density of dark-state
polaritons, which has to be smaller than the atomic density $n$ in order to stay within the linear response limit.  
Since the group velocity $v_{\rm gr}$ can be orders of magnitude larger than the recoil velocity of atoms
$v_{\rm rec}$ and since $k_p L_{\rm abs} \gg 1$, the critical temperature can be much larger than that of the atoms. 
E.g. for $\rho_{\rm DSP}/n = 10^{-1}$, $v_{\rm gr}= 1$ km/s, $v_{\rm rec} = 5$ cm/s, $L_{\rm abs} =$
1 cm, $\Delta/\gamma = 10$ and $k_p = 2\pi/ 500$ nm, yields: $T_c/T_c^{\rm atom} \approx 6 \times 10^5$, i.e. 
a value in the mK regime.  On the other hand the finite frequency window of EIT at large detuning
$k_B T_{\rm EIT} \approx \hbar \Delta$ limits the maximum allowed temperature where losses are
small to the regime of $0.1\dots 1$ mK. The losses outside the EIT transparency window may be used for
evaporative cooling of polaritons, which is however outside the scope of the present paper.

Condensation is achieved differently from atoms or exciton-polaritons.
Instead of changing the temperature or the density one can dynamically change the critical 
temperature of condensation by varying the effective mass, or respectively $\theta$.
Initial preparation of low-energy DSPs can be realized in various ways. Spontaneous Raman scattering cannot be used
since it results in too high polariton momenta on the order of the recoil momentum corresponding to polariton
temperatures on the order of $T_{\rm rec} m/\left(m_\perp^{2/3} m_\parallel^{1/3}\right)$.
Direct RF coupling or storage of a coherent light pulse yield a low energy coherent polariton distribution. 
Alternatively storage of an incoherent light pulse results in the preparation of a non-equilibrium mixed state. 

Since the maximum temperature allowed by the finite transparency window is below the temperature range
where collisions in a gas cell could provide sufficiently fast thermalization, the latter
should occur by elastic collisions between dark-state polaritons. These can be induced 
by resonantly enhanced optical nonlinearities provided by the stationary-light coupling scheme
itself. If e.g.~the excited states in  Fig.\ref{fig1} are hyperfine states
then other off-resonant couplings exist, such as those shown in Fig.\ref{fig1}b.
This coupling induces ac-Stark shifts which give rise to a resonantly enhanced optical Kerr nonlinearity \cite{Schmidt-Imamoglu,Chang-2008}. If we ignore off-resonant couplings of the control laser fields, which is justified if the energy splitting between the lower states $|g\rangle$ and $|s\rangle$ is larger than $|\Omega_\pm|$, the resulting effective Kerr-interaction reads
\begin{eqnarray}
\hat H_{\rm Kerr} = -\hbar \frac{g^2}{\Delta_{\rm Kerr}} \tan^2\theta \int {\rm d}^3\mathbf{r}\, 
: \left( \hat {\cal E}^\dagger_+\hat {\cal E}_+ +\hat {\cal E}^\dagger_- \hat {\cal E}_-\right)^2 :\,\, .
\label{eq:Kerr}
\end{eqnarray}
$\Delta_{\rm Kerr}$ is the one-photon detuning from the additional excited states, and "$:\,\, \, :$" denotes normal ordering.  Since (\ref{eq:Kerr}) only effects the electromagnetic component of the polaritons, the optical Kerr interaction involves
also nonlinear scattering of dark- into bright-state polaritons. Assuming $\sin\theta\approx 1$, i.e. $v_{gr}\ll c$, one finds 
\begin{align}
\hat H_{\text{Kerr}}\,&=-\frac{\hbar g^2 \cos^2\theta}{\Delta_{\text{Kerr}}}\int\text{d}^3\vec r \, \hat\Psi^\dagger\hat\Psi^\dagger\hat\Psi\hat\Psi	\nonumber \\
& +\frac{\hbar g^2 \cos\theta}{\Delta_{\text{Kerr}}}\int\text{d}^3\vec r \, \hat\Psi^\dagger\left(\hat\Phi_2^\dagger\hat\Psi+\hat\Psi^\dagger\hat\Phi_2\right)\hat\Psi.
\end{align}
The bright polaritons decay rapidly and thus the 
nonlinear coupling of polariton modes results in an effective loss. 
Adiabatic elimination of the fast decaying bright polaritons yields in Born-Markov approximation a Liouville equation for the density matrix of the DSPs with a Hamiltonian part describing elastic two-body collisions and a nonlinear loss part. 
\begin{eqnarray}
 \dot\rho_\Psi\,&=\,&-i\frac{g^2\cos^2\theta}{\Delta_{Kerr}}\int\text{d}^3\vec r \, \left[\D^\dagger\D^\dagger\D\D,\rho_\Psi\right] \nonumber\\
  & &\,+4i\frac{g^2\Delta\cos^2\theta}{n\Delta^2_{Kerr}}\int\text{d}^3\vec r \,\left[ \D^\dagger\D^\dagger\D\D^\dagger\D\D,\rho_\Psi \right] \nonumber\\
 & &\,+4\frac{g^2\gamma\cos^2\theta}{n\Delta^2_{Kerr}}\int\text{d}^3\vec r \, \left\{ \D^\dagger\D^\dagger\D\D^\dagger\D\D,\rho_\Psi \right\}_+ \nonumber\\
&&\,-8\frac{g^2\gamma\cos^2\theta}{n\Delta^2_{Kerr}}\int\text{d}^3\vec r\;\D^\dagger\D\D\rho_\Psi\D^\dagger\D^\dagger\D.
\label{eq:lindblad-master-DSP}
\end{eqnarray}
From these one can extract the rate of elastic collisions
\begin{eqnarray}
\Gamma_{\rm coll} = \frac{g^2}{\Delta_{\rm Kerr}} \cos^2\theta\rho_{\rm DSP}
= \frac{v_{\rm gr}}{L_{\rm abs}} \left(\frac{\gamma}{\Delta_{\rm Kerr}} \frac{\rho_{\rm DSP}}{n}\right)
\end{eqnarray}
as well as the rate of collision induced losses
\begin{eqnarray}
\Gamma_{\rm loss}^{\rm nl} = \frac{v_{\rm gr}}{L_{\rm abs}} \, \left(\frac{\gamma}{\Delta_{\rm Kerr}} \frac{\rho_{\rm DSP}}{n}\right)^2.
\end{eqnarray}
One notices that the characteristic time scale $\tau$ of both processes is determined by the ratio 
of absorption length to group velocity. For typical vapor values
such as $v_{\rm gr} = 1$ km/s and $L_{\rm abs} = 1$ cm one finds $\tau = 10^{-5}$ s, while for
a solid or a condensate of atoms where $L_{\rm abs}$ can be as small as 10 $\mu$m one finds $\tau = 10^{-8}$s.
Since $\gamma/\Delta_{\rm Kerr}$ as well as $\rho_{\rm DSP}/n$ are small compared to unity, elastic collisions
happen always on a much shorter time scale than the nonlinear losses.

In addition to the nonlinear loss there is also a linear absorption resulting from the imaginary
part of the polariton mass. It corresponds to the absorption of high spatial-frequency
components in EIT. If $L$ denotes the characteristic longitudinal 
length scale of the polariton, the linear loss rate can be estimated from eq.(\ref{eq:Psi}):
\begin{equation}
\Gamma_{\rm loss}^{\rm lin} = \frac{v_{\rm gr}}{L_{\rm abs}}  \, \left(\frac{L_{\rm abs}}{L}\right)^2.
\end{equation}
In order for the elastic collisions to be fast compared to the linear losses it is necessary that the optical 
depth $OD\equiv L/L_{\rm abs}$ of the
medium over the length $L$ of the polariton wavepacket fulfills
\begin{equation}
\textrm{OD} =
\frac{L}{L_{\rm abs}} > \sqrt{\frac{\Delta_{\rm Kerr}}{\gamma}\frac{n}{\rho_{\rm DSP}}}.
\end{equation}
Taking the above example of $v_{\rm gr}= 1$ km/s and $L_{\rm abs} = 1$ cm or  10 $\mu$m respectively, and 
assuming $\Delta_{\rm Kerr} = 100 \gamma$ and $\rho_{\rm DSP}/n=0.1$ finally yields an
elastic collision rate of $\Gamma_{\rm el} \approx 10^2$  s$^{-1}$, or $10^5$ s$^{-1}$. Thus 
for sufficiently large elastic collision rates one needs a 
large optical depth OD$> 30$, which is feasible however in solid-state systems or magneto-optical traps. 

\begin{figure}[hbt]
    \centering
   \epsfig{file=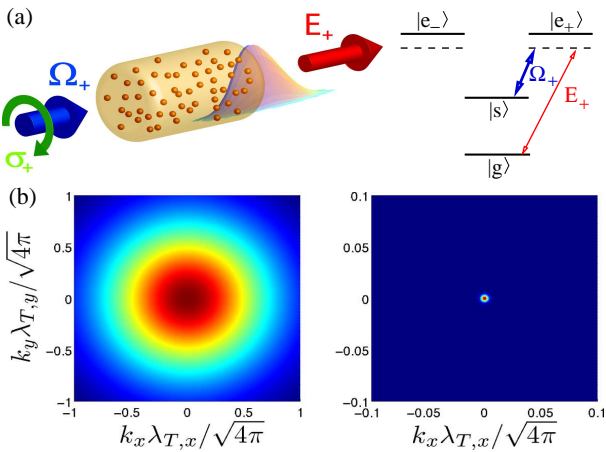, width=8cm}
      \caption{(Color online) {\it top:} detection of condensation: switching off one of the two control beams
transforms the stationary polariton into a propagating polariton which can be observed as 
light pulse emitted from the sample. {\it bottom:} Transverse emission profile above (left) and
below (right) condensation. An ideal quasi-homogeneous Bose-gas was assumed with finite transversal extend
modelled by a Gaussian distribution of width 178 $\lambda_{T,x}$ with $\lambda_{T,x}$ being the
transversal de-Broglie wavelength at temperature $T$. 
}
      \label{fig3}
\end{figure}

The signature of condensation is here the macroscopic occupation of the $\vec k=0$ momentum state. The transition into modes with vanishing transverse momentum $\vec k_\perp=0$  can easily be observed by switching off one of the two control laser. In this case the stationary polariton will be transformed into a moving one and will leave the sample as a light pulse. 
>From the transverse profile of the emitted pulse one can deduce the population of the $\vec k_\perp$  momentum modes
as indicated in Fig. \ref{fig3}.

In summary we have shown that stationary light polaritons behave as massive Schr\"odinger-like particles
with small and variable effective mass. In the presence of a resonantly enhanced optical Kerr nonlinearity
the stationary DSPs undergo elastic collisions at a rate determined by the group
velocity divided by the absorption length multiplied by the inverse normalized Kerr detuning. For a 
sufficiently high optical density of the sample the elastic collisions are fast enough to mediate
a stimulated transition into the $\vec k=0$ polariton mode. The critical temperature for this 
can be orders of magnitude larger than that of atoms. Condensation
can be observed by releasing the stationary field into a propagating field and observing the transverse
mode profile as well as the temporal coherence. In contrast to exciton-polaritons no cavity is needed
and the polariton gas is three dimensional. As a consequence the stimulated transition into the
$\vec k =0$ mode is a true Bose-Einstein condensation. 

\

The authors would like to thank E. S. Demler, G. Morigi and M. Weitz for stimulating discussions.
The financial support of the DFG through the GRK 792 and of the center
of excellence OPTIMAS is gratefully acknowledged.

\

\end{document}